\documentclass[twocolumn,letterpaper,aps,prc,longbibliography,superscriptaddress,nofootinbib,floatfix]{revtex4-2}

\usepackage{graphicx}	   % Include figure files
\usepackage{xspace}        % Include xspace
\usepackage{amsmath}       % Include amsmath

\usepackage{dcolumn}% Align table columns on decimal point
\usepackage{bm}% bold math
\newcommand{\ptpi}{\ensuremath{p_{T,\pi^0}}\xspace}
\newcommand{\pth}{\ensuremath{p_{T,h}}\xspace}

\begin{document}

\title{Jet modification via $\pi^0$-hadron correlations in Au$+$Au 
collisions at $\sqrt{s_{_{NN}}}=200$ GeV}

\newcommand{\abilene}{Abilene Christian University, Abilene, Texas 79699, USA}
\newcommand{\acadsin}{Institute of Physics, Academia Sinica, Taipei 11529, Taiwan}
\newcommand{\augie}{Department of Physics, Augustana University, Sioux Falls, South Dakota 57197, USA}
\newcommand{\banaras}{Department of Physics, Banaras Hindu University, Varanasi 221005, India}
\newcommand{\barc}{Bhabha Atomic Research Centre, Bombay 400 085, India}
\newcommand{\baruch}{Baruch College, City University of New York, New York, New York, 10010 USA}
\newcommand{\bnlcoll}{Collider-Accelerator Department, Brookhaven National Laboratory, Upton, New York 11973-5000, USA}
\newcommand{\bnlphys}{Physics Department, Brookhaven National Laboratory, Upton, New York 11973-5000, USA}
\newcommand{\caucr}{University of California-Riverside, Riverside, California 92521, USA}
\newcommand{\charlesczech}{Charles University, Faculty of Mathematics and Physics, 180 00 Troja, Prague, Czech Republic}
\newcommand{\ciae}{Science and Technology on Nuclear Data Laboratory, China Institute of Atomic Energy, Beijing 102413, People's Republic of China}
\newcommand{\cns}{Center for Nuclear Study, Graduate School of Science, University of Tokyo, 7-3-1 Hongo, Bunkyo, Tokyo 113-0033, Japan}
\newcommand{\colorado}{University of Colorado, Boulder, Colorado 80309, USA}
\newcommand{\columbia}{Columbia University, New York, New York 10027 and Nevis Laboratories, Irvington, New York 10533, USA}
\newcommand{\czechtech}{Czech Technical University, Zikova 4, 166 36 Prague 6, Czech Republic}
\newcommand{\dapnia}{Dapnia, CEA Saclay, F-91191, Gif-sur-Yvette, France}
\newcommand{\debrecen}{Debrecen University, H-4010 Debrecen, Egyetem t{\'e}r 1, Hungary}
\newcommand{\elte}{ELTE, E{\"o}tv{\"o}s Lor{\'a}nd University, H-1117 Budapest, P{\'a}zm{\'a}ny P.~s.~1/A, Hungary}
\newcommand{\ewha}{Ewha Womans University, Seoul 120-750, Korea}
\newcommand{\fit}{Florida Institute of Technology, Melbourne, Florida 32901, USA}
\newcommand{\fsu}{Florida State University, Tallahassee, Florida 32306, USA}
\newcommand{\gsu}{Georgia State University, Atlanta, Georgia 30303, USA}
\newcommand{\hiroshima}{Physics Program and International Institute for Sustainability with Knotted Chiral Meta Matter (SKCM2), Hiroshima University, Higashi-Hiroshima, Hiroshima 739-8526, Japan}
\newcommand{\howard}{Department of Physics and Astronomy, Howard University, Washington, DC 20059, USA}
\newcommand{\hunrenatomki}{HUN-REN ATOMKI, H-4026 Debrecen, Bem t{\'e}r 18/c, Hungary}
\newcommand{\ihepprot}{IHEP Protvino, State Research Center of Russian Federation, Institute for High Energy Physics, Protvino, 142281, Russia}
\newcommand{\illuiuc}{University of Illinois at Urbana-Champaign, Urbana, Illinois 61801, USA}
\newcommand{\inrras}{Institute for Nuclear Research of the Russian Academy of Sciences, prospekt 60-letiya Oktyabrya 7a, Moscow 117312, Russia}
\newcommand{\instpasczech}{Institute of Physics, Academy of Sciences of the Czech Republic, Na Slovance 2, 182 21 Prague 8, Czech Republic}
\newcommand{\isu}{Iowa State University, Ames, Iowa 50011, USA}
\newcommand{\jaea}{Advanced Science Research Center, Japan Atomic Energy Agency, 2-4 Shirakata Shirane, Tokai-mura, Naka-gun, Ibaraki-ken 319-1195, Japan}
\newcommand{\jeonbuk}{Jeonbuk National University, Jeonju, 54896, Korea}
\newcommand{\jinrdubna}{Joint Institute for Nuclear Research, 141980 Dubna, Moscow Region, Russia}
\newcommand{\jyvaskyla}{Helsinki Institute of Physics and University of Jyv{\"a}skyl{\"a}, P.O.Box 35, FI-40014 Jyv{\"a}skyl{\"a}, Finland}
\newcommand{\kek}{KEK, High Energy Accelerator Research Organization, Tsukuba, Ibaraki 305-0801, Japan}
\newcommand{\korea}{Korea University, Seoul 02841, Korea}
\newcommand{\kurchatov}{National Research Center ``Kurchatov Institute", Moscow, 123098 Russia}
\newcommand{\kyoto}{Kyoto University, Kyoto 606-8502, Japan}
\newcommand{\labllr}{Laboratoire Leprince-Ringuet, Ecole Polytechnique, CNRS-IN2P3, Route de Saclay, F-91128, Palaiseau, France}
\newcommand{\lahorelums}{Physics Department, Lahore University of Management Sciences, Lahore 54792, Pakistan}
\newcommand{\lawllnl}{Lawrence Livermore National Laboratory, Livermore, California 94550, USA}
\newcommand{\losalamos}{Los Alamos National Laboratory, Los Alamos, New Mexico 87545, USA}
\newcommand{\lpc}{LPC, Universit{\'e} Blaise Pascal, CNRS-IN2P3, Clermont-Fd, 63177 Aubiere Cedex, France}
\newcommand{\lund}{Department of Physics, Lund University, Box 118, SE-221 00 Lund, Sweden}
\newcommand{\lyon}{IPNL, CNRS/IN2P3, Univ Lyon, Universit{\'e} Lyon 1, F-69622, Villeurbanne, France}
\newcommand{\maryland}{University of Maryland, College Park, Maryland 20742, USA}
\newcommand{\mass}{Department of Physics, University of Massachusetts, Amherst, Massachusetts 01003-9337, USA}
\newcommand{\mate}{MATE, Laboratory of Femtoscopy, K\'aroly R\'obert Campus, H-3200 Gy\"ongy\"os, M\'atrai\'ut 36, Hungary}
\newcommand{\michigan}{Department of Physics, University of Michigan, Ann Arbor, Michigan 48109-1040, USA}
\newcommand{\miss}{Mississippi State University, Mississippi State, Mississippi 39762, USA}
\newcommand{\muenster}{Institut f\"ur Kernphysik, University of M\"unster, D-48149 M\"unster, Germany}
\newcommand{\muhlenberg}{Muhlenberg College, Allentown, Pennsylvania 18104-5586, USA}
\newcommand{\myongji}{Myongji University, Yongin, Kyonggido 449-728, Korea}
\newcommand{\nagasaki}{Nagasaki Institute of Applied Science, Nagasaki-shi, Nagasaki 851-0193, Japan}
\newcommand{\nara}{Nara Women's University, Kita-uoya Nishi-machi Nara 630-8506, Japan}
\newcommand{\natmephi}{National Research Nuclear University, MEPhI, Moscow Engineering Physics Institute, Moscow, 115409, Russia}
\newcommand{\newmex}{University of New Mexico, Albuquerque, New Mexico 87131, USA}
\newcommand{\nmsu}{New Mexico State University, Las Cruces, New Mexico 88003, USA}
\newcommand{\northcg}{Physics and Astronomy Department, University of North Carolina at Greensboro, Greensboro, North Carolina 27412, USA}
\newcommand{\ohio}{Department of Physics and Astronomy, Ohio University, Athens, Ohio 45701, USA}
\newcommand{\ornl}{Oak Ridge National Laboratory, Oak Ridge, Tennessee 37831, USA}
\newcommand{\orsay}{IPN-Orsay, Univ.~Paris-Sud, CNRS/IN2P3, Universit\'e Paris-Saclay, BP1, F-91406, Orsay, France}
\newcommand{\peking}{Peking University, Beijing 100871, People's Republic of China}
\newcommand{\pnpi}{PNPI, Petersburg Nuclear Physics Institute, Gatchina, Leningrad region, 188300, Russia}
\newcommand{\riken}{RIKEN Nishina Center for Accelerator-Based Science, Wako, Saitama 351-0198, Japan}
\newcommand{\rikjrbrc}{RIKEN BNL Research Center, Brookhaven National Laboratory, Upton, New York 11973-5000, USA}
\newcommand{\rikkyo}{Physics Department, Rikkyo University, 3-34-1 Nishi-Ikebukuro, Toshima, Tokyo 171-8501, Japan}
\newcommand{\saispbstu}{Saint Petersburg State Polytechnic University, St.~Petersburg, 195251 Russia}
\newcommand{\saopaulo}{Universidade de S{\~a}o Paulo, Instituto de F\'{\i}sica, Caixa Postal 66318, S{\~a}o Paulo CEP05315-970, Brazil}
\newcommand{\seoulnat}{Department of Physics and Astronomy, Seoul National University, Seoul 151-742, Korea}
\newcommand{\stonybrkc}{Chemistry Department, Stony Brook University, SUNY, Stony Brook, New York 11794-3400, USA}
\newcommand{\stonycrkp}{Department of Physics and Astronomy, Stony Brook University, SUNY, Stony Brook, New York 11794-3800, USA}
\newcommand{\subatech}{SUBATECH (Ecole des Mines de Nantes, CNRS-IN2P3, Universit{\'e} de Nantes) BP 20722-44307, Nantes, France}
\newcommand{\tenn}{University of Tennessee, Knoxville, Tennessee 37996, USA}
\newcommand{\titech}{Department of Physics, Tokyo Institute of Technology, Oh-okayama, Meguro, Tokyo 152-8551, Japan}
\newcommand{\tsukuba}{Tomonaga Center for the History of the Universe, University of Tsukuba, Tsukuba, Ibaraki 305, Japan}
\newcommand{\usmma}{United States Merchant Marine Academy, Kings Point, New York 11024, USA}
\newcommand{\vandy}{Vanderbilt University, Nashville, Tennessee 37235, USA}
\newcommand{\waseda}{Waseda University, Advanced Research Institute for Science and Engineering, 17  Kikui-cho, Shinjuku-ku, Tokyo 162-0044, Japan}
\newcommand{\weizmann}{Weizmann Institute, Rehovot 76100, Israel}
\newcommand{\wigner}{Institute for Particle and Nuclear Physics, Wigner Research Centre for Physics, Hungarian Academy of Sciences (Wigner RCP, RMKI) H-1525 Budapest 114, POBox 49, Budapest, Hungary}
\newcommand{\yonsei}{Yonsei University, IPAP, Seoul 120-749, Korea}
\newcommand{\zagreb}{Department of Physics, Faculty of Science, University of Zagreb, Bijeni\v{c}ka c.~32 HR-10002 Zagreb, Croatia}
\newcommand{\zambia}{Department of Physics, School of Natural Sciences, University of Zambia, Great East Road Campus, Box 32379, Lusaka, Zambia}
\affiliation{\abilene}
\affiliation{\acadsin}
\affiliation{\augie}
\affiliation{\banaras}
\affiliation{\barc}
\affiliation{\baruch}
\affiliation{\bnlcoll}
\affiliation{\bnlphys}
\affiliation{\caucr}
\affiliation{\charlesczech}
\affiliation{\ciae}
\affiliation{\cns}
\affiliation{\colorado}
\affiliation{\columbia}
\affiliation{\czechtech}
\affiliation{\dapnia}
\affiliation{\debrecen}
\affiliation{\elte}
\affiliation{\ewha}
\affiliation{\fit}
\affiliation{\fsu}
\affiliation{\gsu}
\affiliation{\hiroshima}
\affiliation{\howard}
\affiliation{\hunrenatomki}
\affiliation{\ihepprot}
\affiliation{\illuiuc}
\affiliation{\inrras}
\affiliation{\instpasczech}
\affiliation{\isu}
\affiliation{\jaea}
\affiliation{\jeonbuk}
\affiliation{\jinrdubna}
\affiliation{\jyvaskyla}
\affiliation{\kek}
\affiliation{\korea}
\affiliation{\kurchatov}
\affiliation{\kyoto}
\affiliation{\labllr}
\affiliation{\lahorelums}
\affiliation{\lawllnl}
\affiliation{\losalamos}
\affiliation{\lpc}
\affiliation{\lund}
\affiliation{\lyon}
\affiliation{\maryland}
\affiliation{\mass}
\affiliation{\mate}
\affiliation{\michigan}
\affiliation{\miss}
\affiliation{\muenster}
\affiliation{\muhlenberg}
\affiliation{\myongji}
\affiliation{\nagasaki}
\affiliation{\nara}
\affiliation{\natmephi}
\affiliation{\newmex}
\affiliation{\nmsu}
\affiliation{\northcg}
\affiliation{\ohio}
\affiliation{\ornl}
\affiliation{\orsay}
\affiliation{\peking}
\affiliation{\pnpi}
\affiliation{\riken}
\affiliation{\rikjrbrc}
\affiliation{\rikkyo}
\affiliation{\saispbstu}
\affiliation{\saopaulo}
\affiliation{\seoulnat}
\affiliation{\stonybrkc}
\affiliation{\stonycrkp}
\affiliation{\subatech}
\affiliation{\tenn}
\affiliation{\titech}
\affiliation{\tsukuba}
\affiliation{\usmma}
\affiliation{\vandy}
\affiliation{\waseda}
\affiliation{\weizmann}
\affiliation{\wigner}
\affiliation{\yonsei}
\affiliation{\zagreb}
\affiliation{\zambia}
\author{N.J.~Abdulameer} \affiliation{\debrecen} \affiliation{\hunrenatomki}
\author{U.~Acharya} \affiliation{\gsu}
\author{A.~Adare} \affiliation{\colorado} 
\author{S.~Afanasiev} \affiliation{\jinrdubna} 
\author{C.~Aidala} \affiliation{\mass} \affiliation{\michigan} 
\author{N.N.~Ajitanand} \altaffiliation{Deceased} \affiliation{\stonybrkc} 
\author{Y.~Akiba} \email[PHENIX Spokesperson: ]{akiba@rcf.rhic.bnl.gov} \affiliation{\riken} \affiliation{\rikjrbrc}
\author{H.~Al-Bataineh} \affiliation{\nmsu} 
\author{J.~Alexander} \affiliation{\stonybrkc} 
\author{M.~Alfred} \affiliation{\howard} 
% ...
\collaboration{PHENIX Collaboration}  \noaffiliation

%-----------------------------------------------------------------------------|

\begin{abstract}

%\linenumbers

High-momentum two-particle correlations are a useful tool for studying 
jet-quenching effects in the quark-gluon plasma. Angular correlations 
between neutral-pion triggers and charged hadrons with transverse 
momenta in the range 4--12~GeV/$c$ and 0.5--7~GeV/$c$, respectively, 
have been measured by the PHENIX experiment in 2014 for Au$+$Au 
collisions at $\sqrt{s_{_{NN}}}=200$~GeV. Suppression is observed in 
the yield of high-momentum jet fragments opposite the trigger particle, 
which indicates jet suppression stemming from in-medium partonic energy 
loss, while enhancement is observed for low-momentum particles. The 
ratio and differences between the yield in Au$+$Au collisions and 
$p$$+$$p$ collisions, $I_{AA}$ and $\Delta_{AA}$, as a function of the 
trigger-hadron azimuthal separation, $\Delta\phi$, are measured for the 
first time at the Relativistic Heavy Ion Collider. These results better 
quantify how the yield of low-$p_T$ associated hadrons is enhanced at 
wide angle, which is crucial for studying energy loss as well as 
medium-response effects.

\end{abstract}

\maketitle

\section{Introduction}

Jets, collimated sprays of energetic particles originating from the 
fragmentation of hard-scattered partons, are an important probe of the 
quark-gluon plasma (QGP) created in ultra-relativistic collisions of 
heavy ions, such as those at the Relativistic Heavy Ion Collider (RHIC) 
and the Large Hadron Collider (LHC)~\cite{bib:JetReview}. In 
particular, these hard-scattered partons interact with the QGP and lose 
energy when traveling through the medium before fragmenting into 
final-state jet particles. This partonic energy loss gives rise to jets 
that have been modified relative to jets that are measured in $p$$+$$p$ 
collisions, where no QGP medium is formed. The momentum distribution as 
well as the spatial distribution of particles within the resulting jets 
in particular are seen to be modified~\cite{bib:ppg210, 
bib:STARjetquenching2,bib:ALICEjetquenching, bib:ATLASjetquenching, 
bib:CMSjetquenching}. Measurements of jet modification allow for direct 
quantification of the energy transport properties of the medium 
\cite{bib:JETqhat}. Once the parton shower interacts with the QGP, the 
jets and medium particles are intrinsically coupled to one another. 
Therefore, the observed modifications can also embody a response from 
the QGP, which is often referred to as a medium 
response~\cite{bib:CoLBT21, bib:Hybrid14}.

\begin{figure}[htp]
    \includegraphics[width=1.0\linewidth]{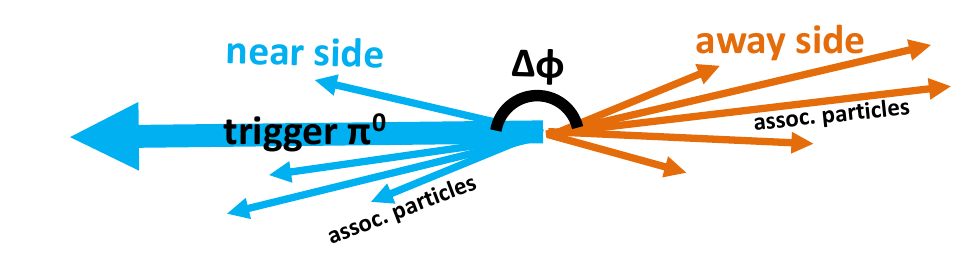}
    \caption{Cartoon of two back-to-back jets as a spray of particles. 
The indicated angle, $\Delta\phi$, measures the azimuthal 
separation between the trigger $\pi^0$ and each associated particle.
The jet labeled ``near side" contains the trigger 
$\pi^0$ at $\Delta\phi$=0. The jet labeled ``away side" shows the
constituents of the recoil jet at $\Delta\phi\approx\pi$. 
}
    \label{fig:jetCartoon}
\end{figure}

High-transverse-momentum neutral pions, $\pi^0$, can be reconstructed 
via their two-photon decay channel and used as jet proxies as they 
carry a large fraction of the jet momentum. Measuring the angular 
correlations between the $\pi^0$ and charged hadrons in the event, 
reveals how charged hadrons are distributed in the jet triggered by the 
$\pi^0$ as well as the opposing jet that appears 180 degrees away from 
the $\pi^0$. This phenomenon is depicted in Fig.~\ref{fig:jetCartoon}. 
The angle, $\Delta\phi$, measures the azimuthal separation between the 
trigger $\pi^0$ and each associated particle. The jet containing the 
trigger $\pi^0$ labeled ``near side" shows the trigger $\pi^0$ itself 
at $\Delta\phi=0$, surrounded by ``near side" associated particles. The 
recoil jet labeled ``away side" shows the associated particles with 
$\Delta\phi\approx\pi$. The abundance of neutral pions, which can be 
reconstructed using the high-granularity PHENIX electromagnetic 
calorimeter (EMCal) out to high $p_T$, are great candidates for trigger 
particles. Two-particle correlations, such as $\pi^0$-hadron 
correlations, are preferred over full-jet reconstruction for dijet 
measurements in PHENIX to overcome the limited PHENIX acceptance.

The previous $\pi^0$-hadron correlations results from 
PHENIX~\cite{bib:PPG106} used an earlier and smaller data set from 
2007. In subtraction of the underlying event, the third- and 
fourth-order harmonics, $v_3$ and $v_4$, were not considered.  
Therefore, the correlations related to jets were not fully decoupled 
from correlations with the underlying event. The 2014 results presented 
here use the largest Au$+$Au data set ever collected by PHENIX and 
include underlying event subtraction using updated measurements of the 
higher-order harmonic terms. The improved statistical precision and 
purity of the measurement enables comparisons of the away-side 
correlation yield in Au$+$Au to that in $p$$+$$p$ as a function of 
$\Delta\phi$, which provides insight into how the distribution of 
particles correlated with the jet is modified.

%----------------------------------------------------- Fig_1

%-------------------------------------------------------------

\section{Experiment}

\begin{figure}[htp]
    \includegraphics[width=1.0\linewidth]{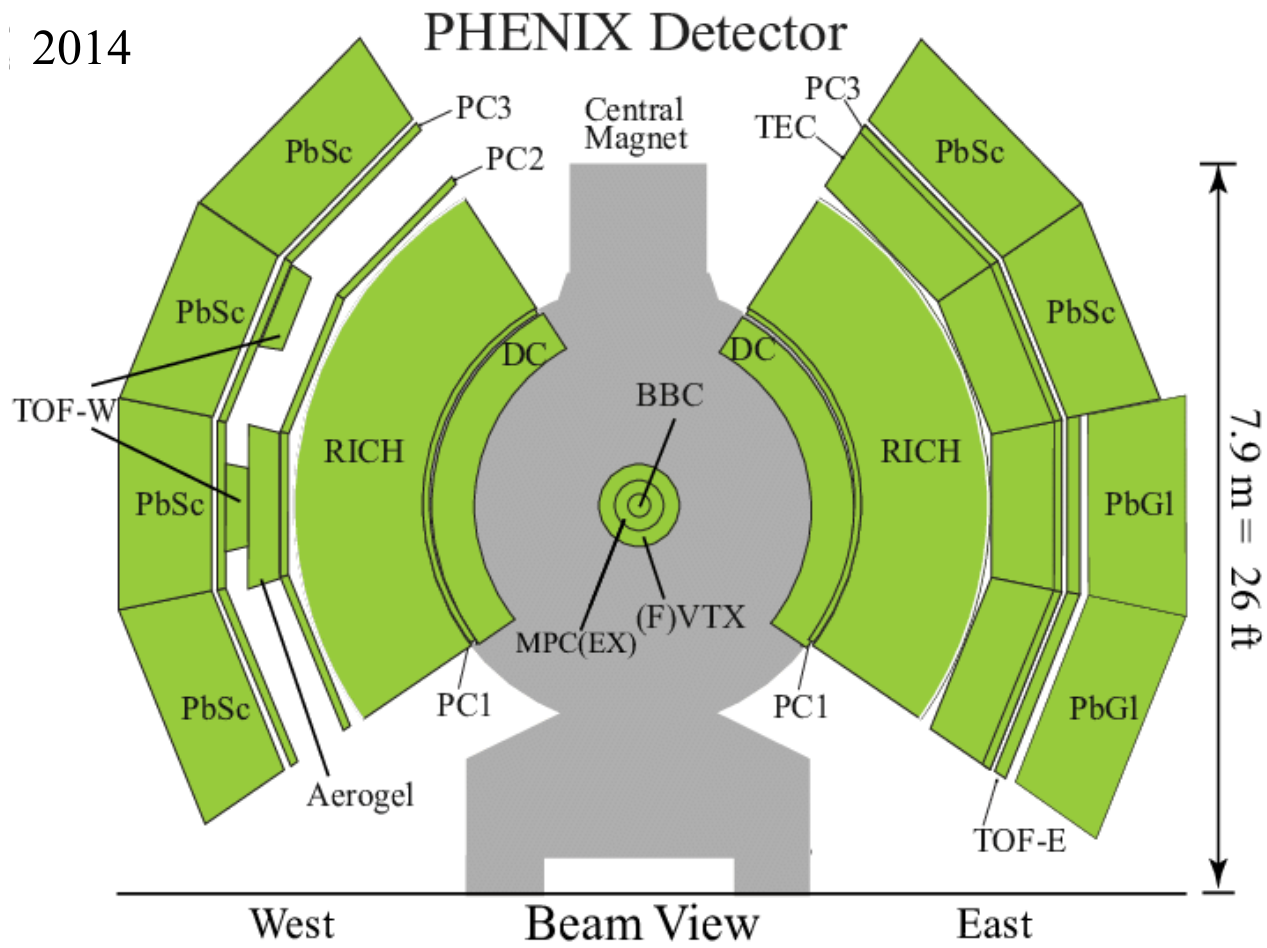}
    \caption{Configuration of PHENIX central arm detector in 2014.}
    \label{fig:PHENIX}
\end{figure}

Figure~\ref{fig:PHENIX} shows the 2014 detector configuration.  In this 
study, the PHENIX collaboration processed 5 billion minimum-bias events 
triggered by the PHENIX beam-beam counters~\cite{bib:PHENIX_inner} and 
collected by the central-arm detectors~\cite{bib:PHENIX_tracking} for 
Au$+$Au collisions at ${\sqrt{s_{_{NN}}}=200}$~GeV. The 
$p$$+$$p$-collision data at ${\sqrt{s_{_{NN}}}=200}$~GeV were collected 
by PHENIX in 2006 and used 3.2~million high-$p_T$ photon-triggered 
events for baseline measurements~\cite{bib:PPG106}.

%----------------------------------------------------- Fig_2

%-------------------------------------------------------------
\section{Data Analysis}

The $\pi^0$'s, which are used as a jet proxy in this analysis, are 
reconstructed from their decay photons by pairing together EMCal 
clusters with an energy of 1~GeV or greater. To remove contamination 
from charged particles, EMCal clusters are required to be greater than 
8~cm away from the closest track projection from the drift chambers to 
the EMCal. Additionally, a cut is made on the cluster shape to remove 
further potential contamination from hadrons. The photon pairs must 
have an energy asymmetry (${\alpha = \frac{|E_{\gamma_1} - 
E_{\gamma_2}|}{E_{\gamma_1} + E_{\gamma_2}}}$, where $E_{\gamma_1}$ and 
$E_{\gamma_2}$ are the energies of the first and second photon, 
respectively) of less than $80\%$ of the sum of the photon energy. 
Finally, each reconstructed $\pi^0$ is required to have an invariant 
mass between $0.12$ and $0.16$~GeV/$c^2$. Reconstructed $\pi^0$'s used 
as jet proxies in this analysis have transverse momenta, \ptpi, of 
4--12~GeV/$c$.

Reconstructed $\pi^0$'s are then paired with reconstructed charged 
tracks. Reconstructed tracks are required to have 
$0.5\leq\pth\leq7$~GeV/$c$, where the upper limit of $7$~GeV/$c$ is 
chosen to limit contamination from secondaries produced by high-$p_T$ 
hadrons within the detector that are misreconstructed as high-$p_T$ 
tracks.

The $\Delta\phi$ correlation functions between $\pi^0$'s and associated charged hadrons are normalized by the number of $\pi^0$'s, $N_{\pi^0}$ and then corrected for the single-hadron reconstruction efficiency, $\epsilon$, and the detector acceptance via simulation and event mixing.  To obtain the correlation functions purely from jets, correlations due to the underlying event and flow are subtracted from the correlation functions. Then, the jet function, which is the differential yield of jet-associated $\pi^0$-hadron pairs per number of $\pi^0$'s in a given $\pi^0$ $p_T$ bin, $N_{\pi^0-h}$, with respect to $\Delta\phi$, can be written as 
\begin{eqnarray}\label{eq:jet}
    \frac{1}{N_{\pi^0}}\frac{dN_{\pi^0-h}}{d\Delta\phi}&=&
    \frac{1}{N_{\pi^0}}\frac{N_{\pi^0-h}}{\epsilon\int{d\Delta\phi}}
    \Bigg\{\frac{dN^{\rm same}_{\pi^0-h}/d\Delta\phi}
{dN^{\rm mix}_{\pi^0-h}/d\Delta\phi} \\
    &-&b_0\left[1+2\sum^4_{n=2}\langle{}v^{\pi^0}_nv^{h}_n\rangle\cos(n\cdot\Delta\phi)\right]\Bigg\} \nonumber 
\end{eqnarray}

\noindent where $N^{\rm same}_{\pi^0-h}$ and $N^{\rm mix}_{\pi^0-h}$ 
are the number of same-event and mixed-event $\pi^0$-hadron pairs, 
respectively.

The contribution to the correlation due to flow appears in the second 
term of Eq.~\eqref{eq:jet} as a Fourier series in terms of the 
azimuthal correlation angle. The coefficient $b_0$ of the Fourier 
series is the magnitude of the underlying event estimated using 
zero-yield-at-minimum method (ZYAM) and absolute background 
normalization method (ABS)~\cite{bib:AbsNorm} in low ${\pth<1}$~GeV/$c$ 
and high ${\pth\geq1}$~GeV/$c$, respectively. To improve the purity of 
the extracted jet-hadron correlation signal, the second to the 
fourth-order harmonics are subtracted (${v_2-v_4}$). The first-order 
harmonic ($v_1$) is not accounted for because its contribution is 
expected to be negligible at 
midrapidity~\cite{bib:STARv1_2008,bib:STARv1_2014}. The 
$n$\textsuperscript{th}-order flow-harmonic coefficients are factorized 
to $v^{\pi^0}_n$ and $v^h_n$ for $\pi^0$'s and charged hadrons, 
respectively.

The $\pi^0$ $v^{\pi^0}_2$ and charged hadron $v^h_n$ in Au$+$Au 
collisions at $200$~GeV come from previous PHENIX 
measurements~\cite{bib:PPG173,bib:PPG110}. However, the higher-order 
$\pi^0$ flow-harmonic coefficients $n=3,4$ in these momentum ranges 
have not been measured at RHIC energies. Thus, to estimate 
$v^{\pi^0}_3$ and $v^{\pi^0}_4$, acoustic 
scaling~\cite{bib:AcousticScaling} is applied. Acoustic scaling is the 
observation that there is a $p_T$-independent relation between 
different powers of the various flow harmonics given by the scaling 
factors, $g_n$, defined as:
\begin{align}\label{eq:gn}
     g_n=\frac{v_n}{(v_2)^{n/2}}.
\end{align}

Assuming the scaling factors of $\pi^0$'s and charged hadron are 
approximately equal due to isospin symmetry (i.e. $g^h_n=g^{\pi^0}_n$), 
$v^{\pi^0}_3$ and $v^{\pi^0}_4$ can then be approximated by rearranging 
Eq.~\eqref{eq:gn} to become:
\begin{align}
     v^{\pi^0}_n =g^h_n\cdot(v^{\pi^0}_2)^{n/2}{\rm  .}
\end{align}

Modification to the per-jet, integrated yield of hadrons is quantified 
by the yield-modification factor $I_{AA}$, defined~as:
\begin{equation}
    I_{AA}(\pth)=\frac{\int^{3\pi/2}_{\pi/2} 
[dN^{\rm AuAu}_{\pi^0-h}/d\Delta\phi]\cdot d\Delta\phi}
{\int^{3\pi/2}_{\pi/2}[dN^{pp}_{\pi^0-h}/d\Delta\phi]\cdot d\Delta\phi}.
\end{equation}

The $I_{AA}$ is defined as the ratio of the integrated per-trigger 
yield of the away-side jet function within 
${\frac{\pi}{2}\leq\Delta\phi\leq\frac{3\pi}{2}}$ in Au$+$Au to that 
measured in $p$$+$$p$ collisions. Additionally, for the first time at 
RHIC, the $I_{AA}$ as a function of $\Delta\phi$, has been measured and 
is defined as the point-by-point ratio of per-trigger yield of the 
away-side jet function in Au$+$Au and $p$$+$$p$, that is,
\begin{align}
    I_{AA}(\Delta\phi)=\frac{dN^{\rm AuAu}_{\pi^0-h}/d\Delta\phi}{dN^{pp}_{\pi^0-h}/d\Delta\phi}{\rm  .}
\end{align}

\noindent Downward fluctuations can cause negative yield at a 
particular $\Delta\phi$ bin. In such cases, the $I_{AA}$ point is not 
shown. Additionally, for clarity, data points with a relative 
statistical or systematic uncertainty equal to or greater than $100$\% 
are also not shown.

Because $I_{AA}(\Delta\phi)$ in regions with small yield in Au$+$Au can 
be inflated through dividing by yields in $p$$+$$p$ close to zero, a 
complimentary observable that can also be extracted is the difference 
between the yields in Au$+$Au and $p$$+$$p$, that is,
\begin{align}
     \Delta_{AA}(\Delta\phi)=\frac{dN^{\rm AuAu}_{\pi^0-h}}{d\Delta\phi}-\frac{dN^{pp}_{\pi^0-h}}{d\Delta\phi}.
\end{align}

%-------------------------------------------------------------%
\section{Systematic Uncertainty}

Seven sources of systematic uncertainty are considered in this 
analysis. The first three arise from the second- to fourth-order 
flow-harmonic coefficients. The fourth is the estimation of the 
underlying event magnitude, $b_0$, using either ZYAM or ABS.  The fifth 
arises from $\pi^0$ reconstruction. The sixth source is the single 
particle efficiency, which is represented by a global scale uncertainty 
of $6.9$\%.  The seventh and final source of systematic uncertainty 
comes from the $p$$+$$p$ measurement used in this analysis, which is 
discussed in detail in Ref.~\cite{bib:PPG106}.

The uncertainties from flow-harmonic coefficients are estimated by 
setting the coefficients to their upper and lower limits individually 
(including the uncertainty of the corresponding scaling factor), 
re-extracting the jet functions, and then re-calculating the observable 
of interest. The relative uncertainties from the flow-harmonic 
coefficients are within a few percent at $\pth>1$~GeV/$c$. Note that, 
the even-order-flow-harmonic coefficients do not contribute to the 
integrated-yield-modification measurements because the integral of the 
even cosine terms equals zero. However, in the lowest $\pth$ bin where 
ZYAM is used in the flow subtraction, $b_0$ is allowed to vary in the 
uncertainties analyses due to flow-harmonic coefficients causing larger 
uncertainty ranges between 10\%--30\% in both differential and 
integrated yield-modification measurements.

The uncertainties arising from $b_0$ itself are estimated by varying 
the $b_0$ obtained from ZYAM and ABS to its upper and lower limits. 
These relative uncertainties are dominant at $\pth<3$~GeV/$c$. The 
relative uncertainties from ABS ranges within 10\% at $\pth>1$~GeV/$c$, 
while the relative uncertainty from ZYAM ranges between 10\%--50\% at 
the lowest $\pth$ bin.

The uncertainty from $\pi^0$ reconstruction is estimated for each 
$\ptpi\otimes\pth$ bin via side-band analysis which involves 
remeasuring the jet functions using photon pairs with an invariant mass 
within 0.65--0.11~GeV/$c^2$ or 0.165--0.2~GeV/$c^2$, instead of the 
nominal $\pi^0$ mass window, 0.12--0.16~GeV/$c^2$. The $\pi^0$ 
reconstruction contribution becomes one of the dominant sources of 
uncertainty as $\pth$ increases. The relative uncertainty from $\pi^0$ 
reconstruction rises from a few percent to 20\%.

Another dominant source of uncertainty at high $\pth$ comes from the 
$p$$+$$p$ collision data.  The relative uncertainty from that increases 
from a few percent at ${2<\pth<3}$~GeV/$c$ to 20\% at 
${5<\pth<7}$~GeV/$c$.

Except the global scaled uncertainty from single particle efficiency, 
uncertainties from other sources are correlated 
data-point-to-data-point. Note that, because the uncertainty from 
$\pi^0$ reconstruction is estimated as a function of $p_T$, it is a 
correlated uncertainty for ${I_{AA}(p_T)}$, but a global scaled 
uncertainty for ${I_{AA}(\Delta\phi)}$ and ${\Delta_{AA}(\Delta\phi)}$.

%-------------------------------------------------------------%
\section{Results}

\begin{figure}[ht]
    \includegraphics[width=1.0\linewidth]{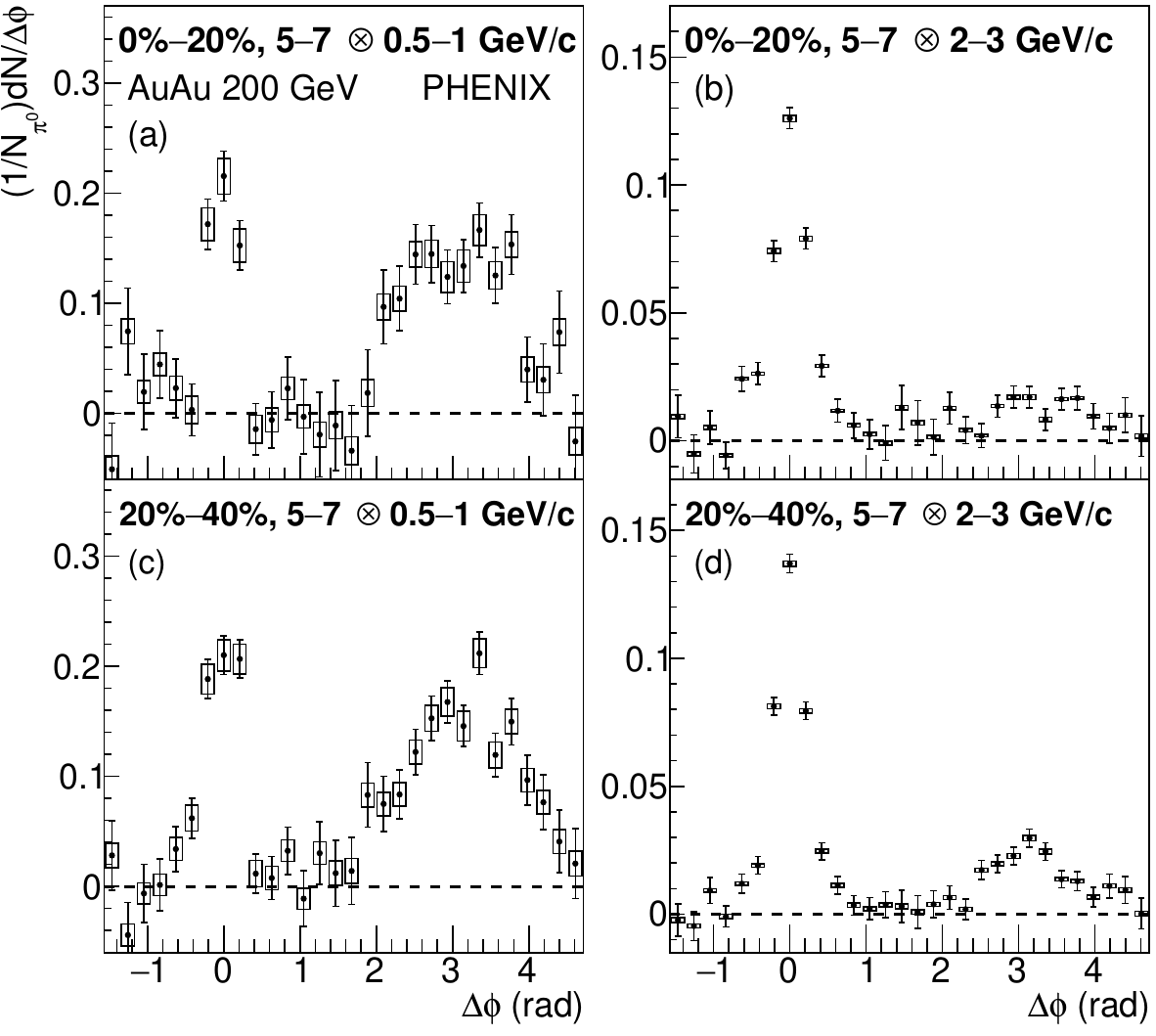}
    \caption{Per-trigger jet-pair yield as a function of $\Delta\phi$ 
for selected $\pi^0$ trigger and charged-hadron-associated $p_T$ 
combinations (${\ptpi\otimes\pth}$) in Au$+$Au collisions. Statistical 
and systematic uncertainties are drawn as vertical lines and boxes, 
respectively. A global scaling uncertainty of $6.9$\% is not shown.}
    \label{fig:jetFunc}
\end{figure}

Figure~\ref{fig:jetFunc} shows the jet functions after subtracting the 
underlying event from the correlation functions in the ${5<\ptpi<7 
\otimes 0.5<\pth<1}$~GeV/$c$ and ${5<\ptpi<7 \otimes 2<\pth<4}$~GeV/$c$ 
momentum bins going left to right, and in the 0\%--20\% and 20\%--40\% 
going from top to bottom. The away-side jet peaks shown in 
Fig.~\ref{fig:jetFunc} appear closer to a Gaussian function compared to 
previous PHENIX results~\cite{bib:PPG106}, where there were pronounced 
peaks appearing to the left and right of the away-side jet peak, a 
phenomenon often attributed to a ``mach-cone" effect created by 
super-sonic traversal of the QGP by hard-scattered partons. However, 
such an effect is no longer seen once contamination from the third and 
fourth harmonics is removed. These changes are more pronounced at low 
\pth where the underlying event is~large.

%----------------------------------------------------- Fig_3

\begin{figure*}[htp]
\includegraphics[width=0.99\linewidth]{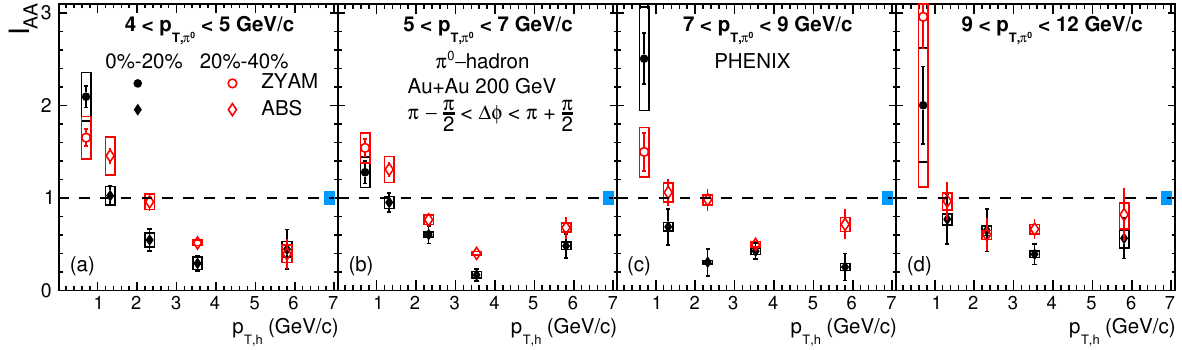}
\caption{\label{fig:IAA_pT} Integrated away-side $I_{AA}$ as a function 
of \pth.  The $\pi^0$ trigger \ptpi range is shown at the top of each 
panel. Statistical and systematic uncertainties are drawn as vertical 
lines and boxes, respectively. A global scaling uncertainty of 6.9\% is 
drawn as a blue box on the right of each panel at $I_{AA}=1$.}
\end{figure*}

The away-side $I_{AA}$ as a function of the associated-hadron momentum, 
${I_{AA}(\pth)}$, is shown in Fig.~\ref{fig:IAA_pT} for four $\pi^0$ 
momentum ranges and in two centrality classes. 

In each $\pi^0$ momentum 
range, the $I_{AA}(\pth)$ is above unity at low $\pth$, but falls as 
$\pth$ increases, eventually reaching below unity at high $\pth$. The 
behavior of the $I_{AA}$ at low-associated hadron momentum indicates 
that there is an enhancement in the yield of soft particles in central 
Au$+$Au collisions, whereas the sub-unity of the $I_{AA}$ at high $p_T$ 
is consistent with a suppression in the yield high-momentum associated 
hadrons. The current understanding of jet-medium interactions indicates 
that in-medium energy loss by high-energy partons is the cause of the 
suppression in the yield of high-momentum hadrons.  However, as shown 
in~\cite{bib:ppg210}, models can reproduce the enhancement measured at 
low momentum by including a mechanism by which energy embedded into the 
medium by hard partons is redistributed into the production of soft 
particles as a medium response.  Unlike in Ref.~\cite{bib:ppg210}, in 
which the $I_{AA}(\pth)$ is measured as a function of $\xi=-ln(z_T)$, 
where $z_T$ is the fraction of $p_T$ carried by the final hadron 
relative to the hard-scattered parton, the transition from enhancement 
to suppression is shown in Fig.~\ref{fig:IAA_pT} to occur at a 
consistent \pth of 1--2~GeV/$c$ in each $\pi^0$ momentum range.  This 
indicates a constant medium response that is independent of the jet 
energy.

%----------------------------------------------------- Fig_4

Lastly, the integrated away-side $I_{AA}$ is measured in the 0\%--20\% 
and 20\%--40\% centrality bins, which are shown in 
Fig.~\ref{fig:IAA_pT} as circle [black] and diamond [red] points, 
respectively. There is no significant centrality dependence observed 
but for ${\pth>2}$~GeV/$c$, the $I_{AA}(\pth)$ in the 20\%--40\% bin is 
systematically closer to unity than in the 0\%--20\% bin. This 
difference in suppression levels could be attributed to a greater 
overall pathlength traversed by hard-scattered partons in the more 
central collisions, which in turn leads to greater energy loss, and a 
lower $I_{AA}(\pth)$ value. This result is qualitatively in agreement 
with results from both the STAR~\cite{bib:STARjetquenching2} and 
ALICE~\cite{bib:alicepi0} collaborations. The difference in the 
magnitude of the enhancement measured by the ALICE experiment (a factor 
of $\approx 5$) vs here (a factor of $\approx 2$) could arise due to 
differences in the plasmas created at the LHC and RHIC, such as the 
mean pathlength traversed by hard partons being larger, leading to an 
increased production of low-$p_T$ hadrons. Similarly, the large 
enhancement measured in this result versus that seen by the STAR 
experiment Ref.~\cite{bib:STARjetquenching2} is due to the fact that 
this measurement extends down to a hadron momentum of 0.5~GeV/$c$, 
where the enhancement is very strong; whereas the threshold is at 
1.2~GeV/$c$ in the STAR result, where the $I_{AA}$ is closer to unity.

%----------------------------------------------------- Fig_5
\begin{figure*}[htp]
\includegraphics[width=0.99\linewidth]{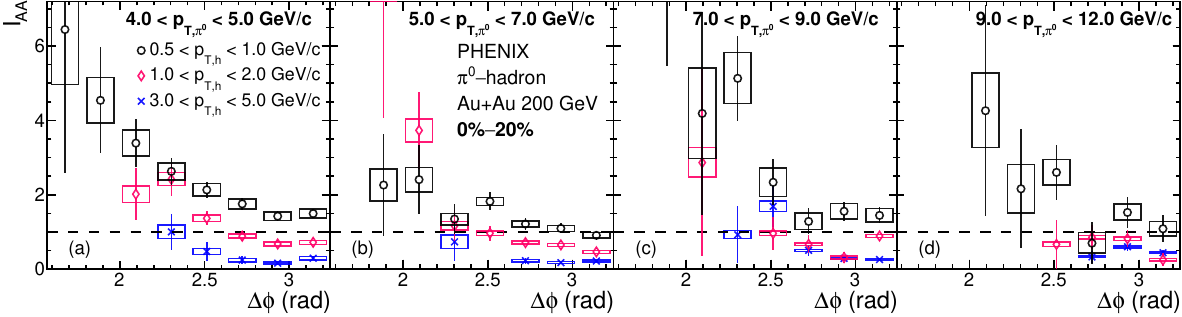}
\includegraphics[width=0.99\linewidth]{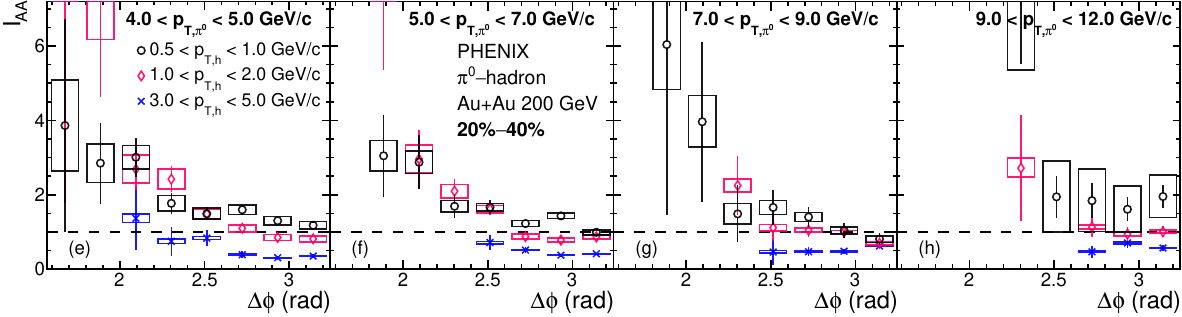}
\caption{\label{fig:IAA_deltaPhi} Differential away-side $I_{AA}$ as a function of $\Delta\phi$ in (a) to (d) 0\%--20\% and (e) to (h) 20\%--40\% centrality classes. The $\pi^0$ trigger \ptpi range is shown at the top of each panel. Statistical and systematic uncertainties are drawn as vertical lines and boxes, respectively. A global uncertainty of 6.9\% is not shown.}
\end{figure*}

Figure~\ref{fig:IAA_deltaPhi} shows the $I_{AA}$ as a function of 
$\Delta\phi$, $I_{AA}(\Delta\phi)$, for three \pth ranges, four \ptpi 
ranges, and two centrality classes. This observable allows for 
quantification of the modification to the jet yield at different 
distances from the away-side jet axis (${\Delta\phi \approx \pi}$). The 
${I_{AA}(\Delta\phi)}$ shows an enhancement in the yield of 
low-momentum hadrons across the away-side jet peak, although this 
enhancement is strongest at wide angles relative to the peak. The 
away-side peak is also the first region where the 
${I_{AA}(\Delta\phi)}$ begins to fall beneath unity as shown by the 
${1.0\leq\pth<2.0}$~GeV/$c$ (red diamonds) in both the 0\%--20\% and 
20\%--40\% centrality bins. In the highest momentum bin reported, 
$3.0\leq\pth<5.0$~GeV/$c$, the yield of charged hadrons is suppressed 
across all angles shown, a result of the partonic energy loss induced 
by parton-medium interactions. In contrast, the enhancement is most 
severe at wide angles relative to the away-side jet peak similar to 
what is seen in Ref.~\cite{bib:ppg210}.

\begin{figure*}[htp]
     \includegraphics[width=0.99\linewidth]{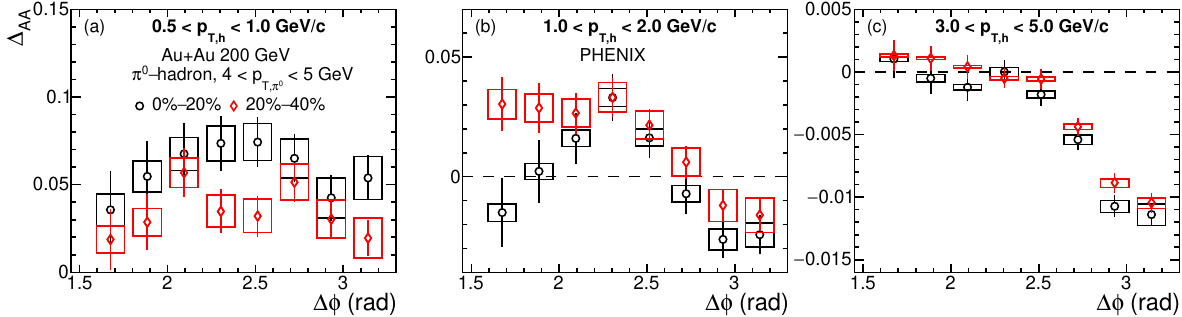}
     \includegraphics[width=0.99\linewidth]{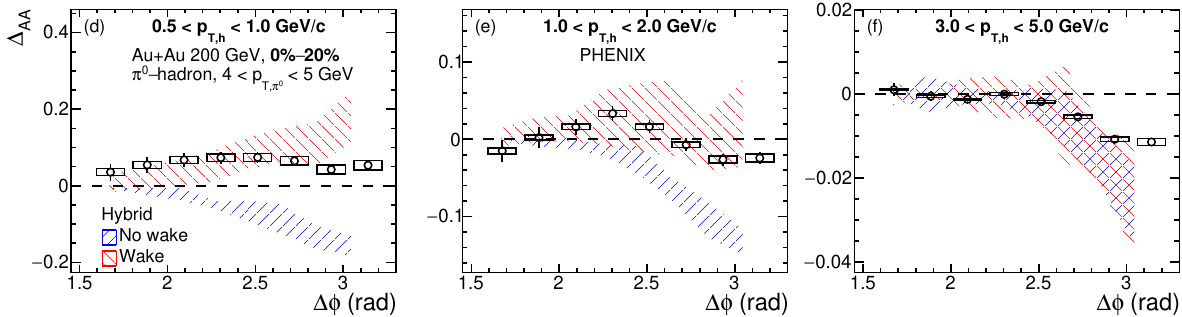}
    \caption{\label{fig:DAA_deltaPhi} (a)--(c): Differential away-side 
$\Delta_{AA}$ in 0\%--20\% (circles [black]) and 20\%--40\% (diamonds 
[red]) centrality classes for ${\pi/2<\Delta\phi<\pi}$. (d)--(f): 
Differential away-side $\Delta_{AA}$ in 0\%--20\% centrality class for 
the same $\Delta\phi$ range compared to hybrid models with ``Wake" 
(backward [red] slashes) and ``No wake" (forward [blue] slashes). A 
global uncertainty of $6.9$\% is not shown.}
\end{figure*}

Figure~\ref{fig:DAA_deltaPhi} shows the difference between Au$+$Au and 
$p$$+$$p$ in the per-trigger yield, $\Delta_{AA}$, as a function of 
$\Delta\phi$ for hadrons with $0.5<p_T<1$~GeV/$c$. The enhancement 
(where the difference between the Au$+$Au and $p$$+$$p$ yields is 
positive) is again observed over a wide range of angles.  The 
enhancement increases when moving away from the away-side jet axis, 
that is ${\Delta\phi=\pi}$. The enhancement seen at wider angles is 
also consistent with the phenomena of jet broadening. It is notable 
that the enhancement is observed near the $\Delta\phi=\pi/2$ region 
because, as shown in Fig.~\ref{fig:jetFunc}, that is the minimum of the 
per-trigger jet-pair yield. One key advantage of taking the difference 
in Au$+$Au and $p$$+$$p$ over computing the $I_{AA}$ is that it is less 
sensitive than the $I_{AA}$ to the $p$$+$$p$ yields fluctuating close 
to zero, particularly near $\Delta\phi=\pi/2$. This approach provides 
stronger constraints on theoretical models than the $I_{AA}$ in these 
regions. The modification seen in Fig.~\ref{fig:DAA_deltaPhi} is 
further explored by observing how the measurement changes as a function 
of hadron $p_T$.

%----------------------------------------------------- Fig_6

Figure~\ref{fig:DAA_deltaPhi} shows the difference in the per-trigger 
yields between Au$+$Au and $p$$+$$p$ as a function of $\Delta\phi$ for 
different $\pth$ bins associated with 4--5~GeV/$c$ $\pi^0$, which 
clearly demonstrates the transition from enhancement at low $\pth$ to 
suppression at high $\pth$. In particular, the suppression in the 
per-trigger yield is most severe near the jet axis 
(${\Delta\phi\approx\pi}$). This suppression pattern differs slightly 
from that seen in measurements at the LHC, such as 
in~\cite{bib:AtlasRadialModification}, where the yield of hadrons 
within a jet is found to be almost unmodified at the jet axis, 
regardless of the momentum range. However, for these RHIC results the 
$I_{AA}$ and $\Delta_{AA}$ vs $\Delta\phi$ are measured from the recoil 
jet opposite the jet containing the trigger $\pi^0$, which imposes 
almost no bias on the recoil jet. Note that anti-$k_T$ jets like those 
measured in Ref.~\cite{bib:AtlasRadialModification} have more stringent 
requirements and could bias the sample of reconstructed jets in Au$+$Au 
to be more similar to those in $p$$+$$p$ collisions.

\begin{figure*}[h!tp]
    \includegraphics[width=0.99\linewidth]{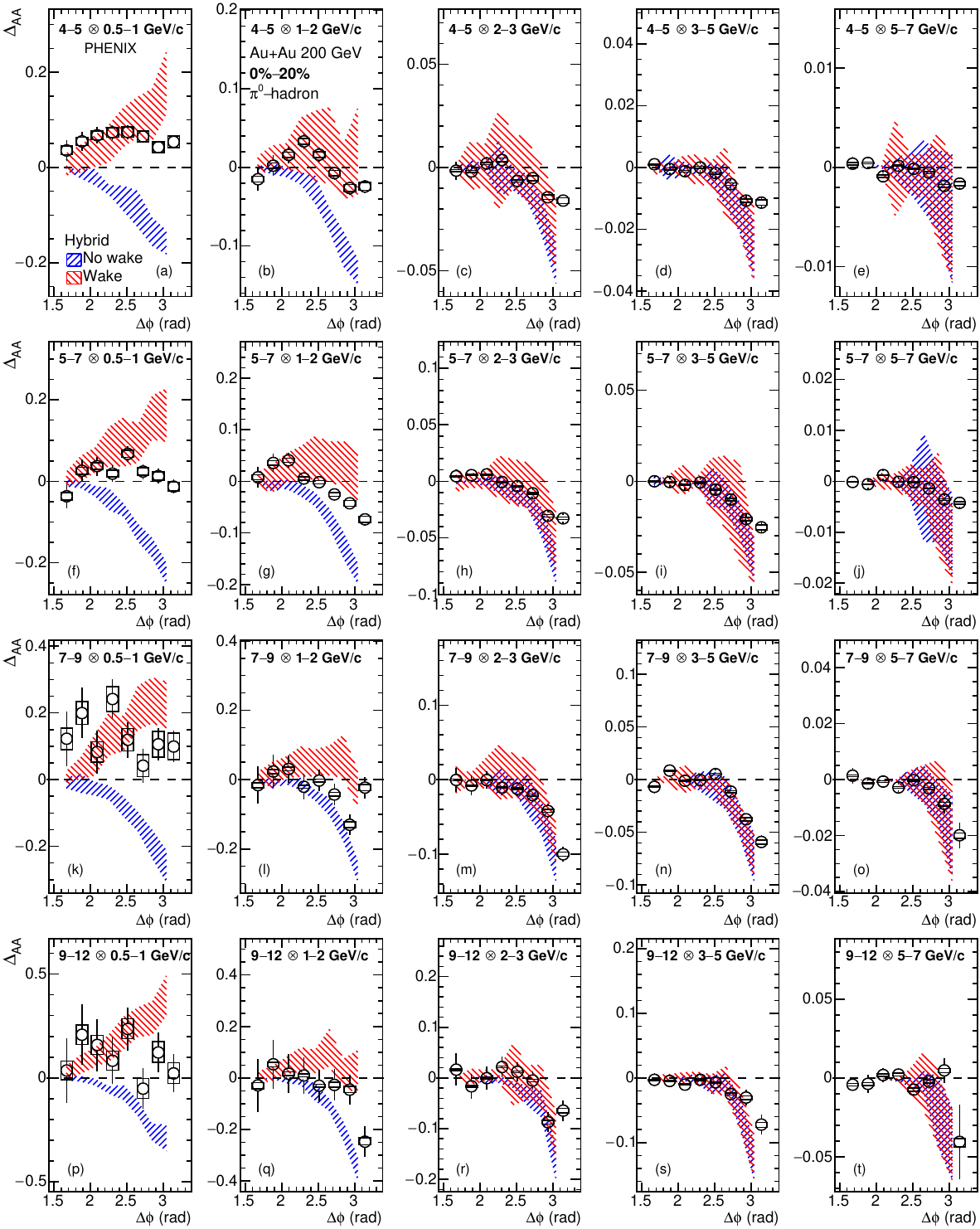}
    \caption{Differential away-side $\Delta_{AA}$ in 0\%--20\% 
centrality for ${\pi/2<\Delta\phi<\pi}$ for various $\pi^0$ trigger and 
charged-hadron-associated $p_T$ combinations ($\ptpi\otimes\pth$). As 
in Fig.~\protect\ref{fig:DAA_deltaPhi}(d)--(f), the ``Wake" and ``No 
wake" hybrid models are overlaid as backward [red] slashes and forward 
[blue] slashes.}
    \label{fig:DAA_C0_allbins}
\end{figure*}

\begin{figure*}[h!tp]
    \includegraphics[width=0.99\linewidth]{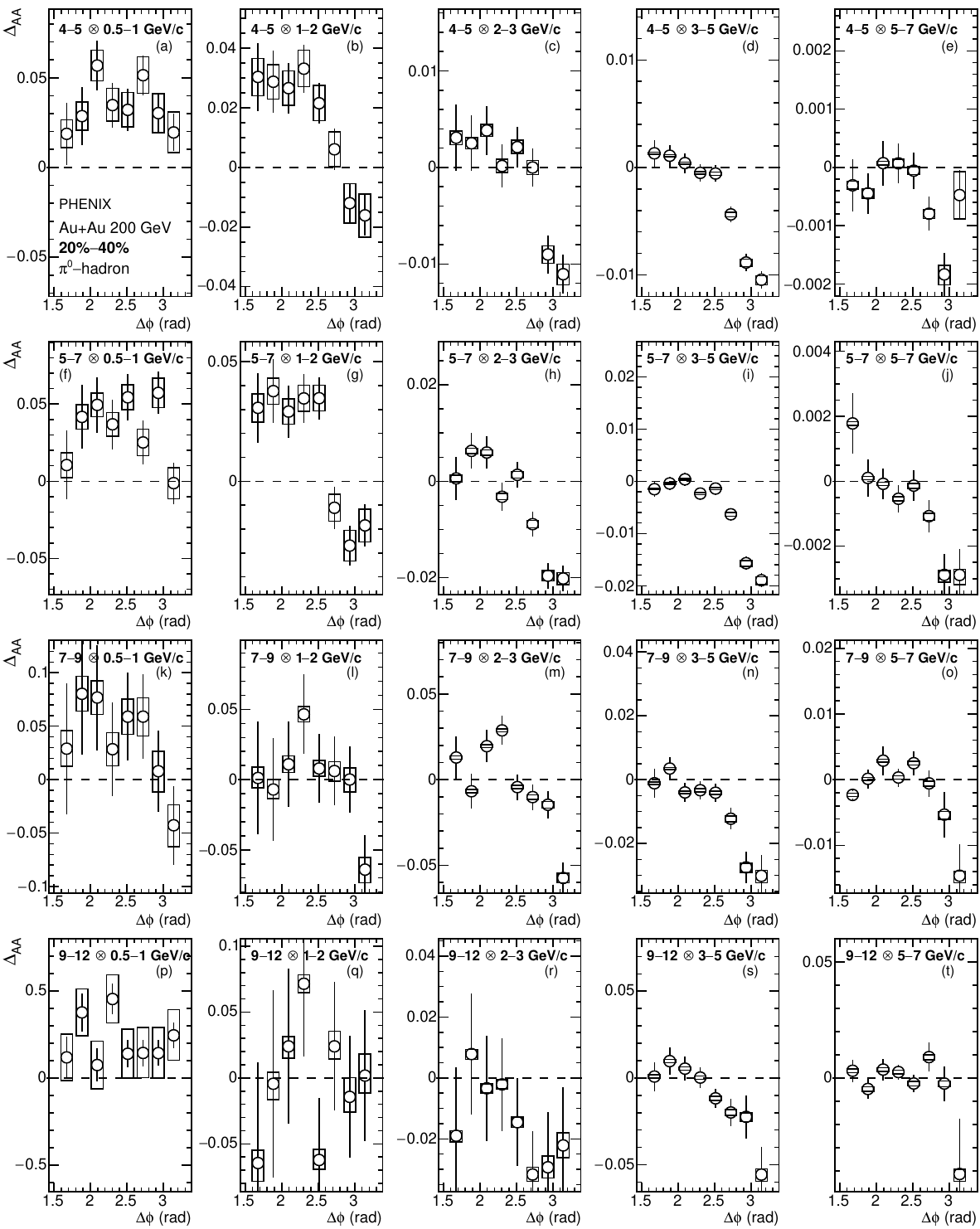}
    \caption{Differential away-side $\Delta_{AA}$ as a function of 
$\Delta\phi$ in 20\%--40\% centrality for various $\pi^0$ trigger and 
charged-hadron-associated $p_T$ combinations ($\ptpi\otimes\pth$).}
    \label{fig:DAA_C1_allbins}
\end{figure*}

Figure~\ref{fig:DAA_deltaPhi} plots (d) to (f) show the Au$+$Au and 
$p$$+$$p$ yield differences versus $\Delta\phi$ for selected 
${\ptpi\otimes\pth}$ bins overlaid with calculations from the HYBRID 
model~\cite{bib:Hybrid14} (all available ${\ptpi\otimes\pth}$ bins are 
shown in Figs.~\ref{fig:DAA_C0_allbins} and \ref{fig:DAA_C1_allbins}). 
This model uses a combination of perturbative quantum chromodynamics 
and anti-de Sitter/conformal field theory to handle hard and soft 
interactions within the medium, respectively. One can see that at high 
$\pth$, the HYBRID model reproduces the data well within the 
uncertainty of the model. Two versions of the model are presented, 
differentiated by how they handle the medium response to the embedded 
partonic energy by the hard-scattered parton. The curve labeled ``Wake" 
models a medium response to the lost energy as a hydrodynamic wake of 
soft particles, which well reproduces the wide-angle enhancement seen 
in the data at low $\pth$. The curve labeled ``No wake" does not 
include this effect, and, thus, fails to reproduce the data at low 
$\pth$. The success of this model at low $\pth$ relies on a 
qualitatively similar mechanism as the CoLBT-Hydro model shown in 
Ref.~\cite{bib:ppg210}.  Both models include hydrodynamic responses 
from the medium that contribute to the creation of an excess of soft 
particles in the final-state particle distribution.

%----------------------------------------------------- Fig_7

%----------------------------------------------------- Fig_8

%-------------------------------------------------------------
\section{Summary}

The PHENIX collaboration presented a new $\pi^0$-hadron correlation 
measurement in Au$+$Au collision at $200$~GeV with data taken in 2014 
at RHIC.  With the enhanced statistics of the 2014 data set and 
improved background subtraction that accounts for contributions from 
flow up to the fourth-order flow coefficient, the results presented 
here are an improvement over previous PHENIX measurements. These jet 
functions and their integrated yields are then used to calculate both 
the quotient, $I_{AA}$, and the difference, $\Delta_{AA}$, between 
Au$+$Au and $p$$+$$p$ yields vs $\Delta\phi$ (as well as the $I_{AA}$) 
as a function of the associated-hadron $p_T$.

The integrated per-trigger-yield modification, $I_{AA}$ as a function 
of \pth, is indicative of partonic energy loss by hard partons via 
parton-medium interactions, leading to the suppression of hard jet 
particles and enhancement of soft jet particles. The new observables, 
differential per-trigger-yield modifications as a function of 
$\Delta\phi$, show the modifications are angularly dependent within the 
recoil jets. The angular dependence of $I_{AA}$ and $\Delta_{AA}$, also 
changes with jet-particle transverse momentum. The transition from 
enhancement of low-momentum particles to suppression at higher momentum 
is consistent with models such as the Hybrid model that include medium 
response. The differential $I_{AA}$ is sensitive to the small 
modification at the edge of the jets, while the differential 
$\Delta_{AA}$ is less sensitive to statistical fluctuations. Using a 
variety of jet related observables will further constrain the models in 
the study of jet modifications, allowing for a more precise 
determination of QGP properties.
 
%%%%%%%%%%%%%%%%%%%%%%  ACKNOWLEDGMENTS}  %%%%% MGS24a0614 version
% to include Run06 countries that were dropped after Run14
%% 2018 change in Korea
%%% 2021 change in dropping Brazil, Germany, and Pakistan, because
%%%      they no longer have active MGS and left PHENIX before 2015
%% 2024 add HUN-REN ATOMKI [and remove some] (Hungary)

\begin{acknowledgments}

We thank the staff of the Collider-Accelerator and Physics
Departments at Brookhaven National Laboratory and the staff of
the other PHENIX participating institutions for their vital
contributions.  
We acknowledge support from the Office of Nuclear Physics in the
Office of Science of the Department of Energy,
the National Science Foundation,
a sponsored research grant from Renaissance Technologies LLC,
Abilene Christian University Research Council,
Research Foundation of SUNY, and
Dean of the College of Arts and Sciences, Vanderbilt University
(U.S.A),
Ministry of Education, Culture, Sports, Science, and Technology
and the Japan Society for the Promotion of Science (Japan),
Conselho Nacional de Desenvolvimento Cient\'{\i}fico e
Tecnol{\'o}gico and Funda\c c{\~a}o de Amparo {\`a} Pesquisa do
Estado de S{\~a}o Paulo (Brazil),
Natural Science Foundation of China (People's Republic of China),
Croatian Science Foundation and
Ministry of Science and Education (Croatia),
Ministry of Education, Youth and Sports (Czech Republic),
Centre National de la Recherche Scientifique, Commissariat
{\`a} l'{\'E}nergie Atomique, and Institut National de Physique
Nucl{\'e}aire et de Physique des Particules (France),
Bundesministerium f\"ur Bildung und Forschung, Deutscher
Akademischer Austausch Dienst, and Alexander von Humboldt Stiftung (Germany),
J. Bolyai Research Scholarship, EFOP, HUN-REN ATOMKI, NKFIH,
and OTKA (Hungary), 
Department of Atomic Energy and Department of Science and Technology (India),
Israel Science Foundation (Israel),
Basic Science Research and SRC(CENuM) Programs through NRF
funded by the Ministry of Education and the Ministry of
Science and ICT (Korea).
Physics Department, Lahore University of Management Sciences (Pakistan),
Ministry of Education and Science, Russian Academy of Sciences,
Federal Agency of Atomic Energy (Russia),
VR and Wallenberg Foundation (Sweden),
University of Zambia, the Government of the Republic of Zambia (Zambia),
the U.S. Civilian Research and Development Foundation for the
Independent States of the Former Soviet Union,
the Hungarian American Enterprise Scholarship Fund,
the US-Hungarian Fulbright Foundation,
and the US-Israel Binational Science Foundation.

\end{acknowledgments}

%\clearpage

%\bibliography{ppg250x1} 

\begin{thebibliography}{20}%
\makeatletter
\providecommand \@ifxundefined [1]{%
 \@ifx{#1\undefined}
}%
\providecommand \@ifnum [1]{%
 \ifnum #1\expandafter \@firstoftwo
 \else \expandafter \@secondoftwo
 \fi
}%
\providecommand \@ifx [1]{%
 \ifx #1\expandafter \@firstoftwo
 \else \expandafter \@secondoftwo
 \fi
}%
\providecommand \natexlab [1]{#1}%
\providecommand \enquote  [1]{``#1''}%
\providecommand \bibnamefont  [1]{#1}%
\providecommand \bibfnamefont [1]{#1}%
\providecommand \citenamefont [1]{#1}%
\providecommand \href@noop [0]{\@secondoftwo}%
\providecommand \href [0]{\begingroup \@sanitize@url \@href}%
\providecommand \@href[1]{\@@startlink{#1}\@@href}%
\providecommand \@@href[1]{\endgroup#1\@@endlink}%
\providecommand \@sanitize@url [0]{\catcode `\\12\catcode `\$12\catcode
  `\&12\catcode `\#12\catcode `\^12\catcode `\_12\catcode `\%12\relax}%
\providecommand \@@startlink[1]{}%
\providecommand \@@endlink[0]{}%
\providecommand \url  [0]{\begingroup\@sanitize@url \@url }%
\providecommand \@url [1]{\endgroup\@href {#1}{\urlprefix }}%
\providecommand \urlprefix  [0]{URL }%
\providecommand \Eprint [0]{\href }%
\providecommand \doibase [0]{https://doi.org/}%
\providecommand \selectlanguage [0]{\@gobble}%
\providecommand \bibinfo  [0]{\@secondoftwo}%
\providecommand \bibfield  [0]{\@secondoftwo}%
\providecommand \translation [1]{[#1]}%
\providecommand \BibitemOpen [0]{}%
\providecommand \bibitemStop [0]{}%
\providecommand \bibitemNoStop [0]{.\EOS\space}%
\providecommand \EOS [0]{\spacefactor3000\relax}%
\providecommand \BibitemShut  [1]{\csname bibitem#1\endcsname}%
\let\auto@bib@innerbib\@empty
%</preamble>
\bibitem [{\citenamefont {Connors}\ \emph {et~al.}(2018)\citenamefont
  {Connors}, \citenamefont {Nattrass}, \citenamefont {Reed},\ and\
  \citenamefont {Salur}}]{bib:JetReview}%
  \BibitemOpen
  \bibfield  {author} {\bibinfo {author} {\bibfnamefont {M.}~\bibnamefont
  {Connors}}, \bibinfo {author} {\bibfnamefont {C.}~\bibnamefont {Nattrass}},
  \bibinfo {author} {\bibfnamefont {R.}~\bibnamefont {Reed}},\ and\ \bibinfo
  {author} {\bibfnamefont {S.}~\bibnamefont {Salur}},\ }\bibfield  {title}
  {\bibinfo {title} {{Jet measurements in heavy ion physics}},\ }\href@noop {}
  {\bibfield  {journal} {\bibinfo  {journal} {Rev. Mod. Phys.}\ }\textbf
  {\bibinfo {volume} {90}},\ \bibinfo {pages} {025005} (\bibinfo {year}
  {2018})}\BibitemShut {NoStop}%
\bibitem [{\citenamefont {Acharya}\ \emph {et~al.}(2020)\citenamefont {Acharya}
  \emph {et~al.}}]{bib:ppg210}%
  \BibitemOpen
  \bibfield  {author} {\bibinfo {author} {\bibfnamefont {U.}~\bibnamefont
  {Acharya}} \emph {et~al.} (\bibinfo {collaboration} {PHENIX Collaboration}),\
  }\bibfield  {title} {\bibinfo {title} {{Measurement of jet-medium
  interactions via direct photon-hadron correlations in Au$+$Au and $d$$+$Au
  collisions at $\sqrt{s_{NN}}=200$ GeV}},\ }\href
  {https://doi.org/10.1103/PhysRevC.102.054910} {\bibfield  {journal} {\bibinfo
   {journal} {Phys. Rev. C}\ }\textbf {\bibinfo {volume} {102}},\ \bibinfo
  {pages} {054910} (\bibinfo {year} {2020})}\BibitemShut {NoStop}%
\bibitem [{\citenamefont {Adamczyk}\ \emph {et~al.}(2016)\citenamefont
  {Adamczyk} \emph {et~al.}}]{bib:STARjetquenching2}%
  \BibitemOpen
  \bibfield  {author} {\bibinfo {author} {\bibfnamefont {L.}~\bibnamefont
  {Adamczyk}} \emph {et~al.} (\bibinfo {collaboration} {STAR Collaboration}),\
  }\bibfield  {title} {\bibinfo {title} {{Jet-like correlations with
  direct-photon and neutral-pion triggers at $\sqrt{s_{NN}}=200$ GeV}},\
  }\href@noop {} {\bibfield  {journal} {\bibinfo  {journal} {Phys. Lett. B}\
  }\textbf {\bibinfo {volume} {760}},\ \bibinfo {pages} {689} (\bibinfo {year}
  {2016})}\BibitemShut {NoStop}%
\bibitem [{\citenamefont {Adam}\ \emph {et~al.}(2015)\citenamefont {Adam} \emph
  {et~al.}}]{bib:ALICEjetquenching}%
  \BibitemOpen
  \bibfield  {author} {\bibinfo {author} {\bibfnamefont {J.}~\bibnamefont
  {Adam}} \emph {et~al.} (\bibinfo {collaboration} {ALICE Collaboration}),\
  }\bibfield  {title} {\bibinfo {title} {{Measurement of jet suppression in
  central {Pb-Pb} collisions at $\sqrt{s_{\rm NN}}$=2.76~{TeV}}},\ }\href@noop
  {} {\bibfield  {journal} {\bibinfo  {journal} {Phys. Lett. B}\ }\textbf
  {\bibinfo {volume} {746}},\ \bibinfo {pages} {1} (\bibinfo {year}
  {2015})}\BibitemShut {NoStop}%
\bibitem [{\citenamefont {Aad}\ \emph {et~al.}(2010)\citenamefont {Aad} \emph
  {et~al.}}]{bib:ATLASjetquenching}%
  \BibitemOpen
  \bibfield  {author} {\bibinfo {author} {\bibfnamefont {G.}~\bibnamefont
  {Aad}} \emph {et~al.} (\bibinfo {collaboration} {ATLAS Collaboration}),\
  }\bibfield  {title} {\bibinfo {title} {{Observation of a Centrality-Dependent
  Dijet Asymmetry in Lead-Lead Collisions at $\sqrt{s_{NN}}=2.77$ TeV with the
  ATLAS Detector at the LHC}},\ }\href@noop {} {\bibfield  {journal} {\bibinfo
  {journal} {Phys. Rev. Lett.}\ }\textbf {\bibinfo {volume} {105}},\ \bibinfo
  {pages} {252303} (\bibinfo {year} {2010})}\BibitemShut {NoStop}%
\bibitem [{\citenamefont {Chatrchyan}\ \emph {et~al.}(2011)\citenamefont
  {Chatrchyan} \emph {et~al.}}]{bib:CMSjetquenching}%
  \BibitemOpen
  \bibfield  {author} {\bibinfo {author} {\bibfnamefont {S.}~\bibnamefont
  {Chatrchyan}} \emph {et~al.} (\bibinfo {collaboration} {CMS Collaboration}),\
  }\bibfield  {title} {\bibinfo {title} {{Observation and studies of jet
  quenching in {PbPb} collisions at nucleon-nucleon center-of-mass energy =
  2.76 TeV}},\ }\href@noop {} {\bibfield  {journal} {\bibinfo  {journal} {Phys.
  Rev. C}\ }\textbf {\bibinfo {volume} {84}},\ \bibinfo {pages} {024906}
  (\bibinfo {year} {2011})}\BibitemShut {NoStop}%
\bibitem [{\citenamefont {Burke}\ \emph {et~al.}(2014)\citenamefont {Burke},
  \citenamefont {Buzzatti}, \citenamefont {Chang}, \citenamefont {Gale},
  \citenamefont {Gyulassy}, \citenamefont {Heinz}, \citenamefont {Jeon},
  \citenamefont {Majumder}, \citenamefont {M{\"u}ller}, \citenamefont {Qin},
  \citenamefont {Schenke}, \citenamefont {Shen}, \citenamefont {Wang},
  \citenamefont {Xu}, \citenamefont {Young},\ and\ \citenamefont
  {Zhang}}]{bib:JETqhat}%
  \BibitemOpen
  \bibfield  {author} {\bibinfo {author} {\bibfnamefont {K.~M.}\ \bibnamefont
  {Burke}}, \bibinfo {author} {\bibfnamefont {A.}~\bibnamefont {Buzzatti}},
  \bibinfo {author} {\bibfnamefont {N.}~\bibnamefont {Chang}}, \bibinfo
  {author} {\bibfnamefont {C.}~\bibnamefont {Gale}}, \bibinfo {author}
  {\bibfnamefont {M.}~\bibnamefont {Gyulassy}}, \bibinfo {author}
  {\bibfnamefont {U.}~\bibnamefont {Heinz}}, \bibinfo {author} {\bibfnamefont
  {S.}~\bibnamefont {Jeon}}, \bibinfo {author} {\bibfnamefont {A.}~\bibnamefont
  {Majumder}}, \bibinfo {author} {\bibfnamefont {B.}~\bibnamefont
  {M{\"u}ller}}, \bibinfo {author} {\bibfnamefont {G.-Y.}\ \bibnamefont {Qin}},
  \bibinfo {author} {\bibfnamefont {B.}~\bibnamefont {Schenke}}, \bibinfo
  {author} {\bibfnamefont {C.}~\bibnamefont {Shen}}, \bibinfo {author}
  {\bibfnamefont {X.-N.}\ \bibnamefont {Wang}}, \bibinfo {author}
  {\bibfnamefont {J.}~\bibnamefont {Xu}}, \bibinfo {author} {\bibfnamefont
  {C.}~\bibnamefont {Young}},\ and\ \bibinfo {author} {\bibfnamefont
  {H.}~\bibnamefont {Zhang}} (\bibinfo {collaboration} {JET Collaboration}),\
  }\bibfield  {title} {\bibinfo {title} {{Extracting the jet transport
  coefficient from jet quenching in high-energy heavy-ion collisions}},\
  }\href@noop {} {\bibfield  {journal} {\bibinfo  {journal} {Phys. Rev. C}\
  }\textbf {\bibinfo {volume} {90}},\ \bibinfo {pages} {014909} (\bibinfo
  {year} {2014})}\BibitemShut {NoStop}%
\bibitem [{\citenamefont {Chen}\ \emph {et~al.}(2021)\citenamefont {Chen},
  \citenamefont {Yang}, \citenamefont {He}, \citenamefont {Ke}, \citenamefont
  {Pang},\ and\ \citenamefont {Wang}}]{bib:CoLBT21}%
  \BibitemOpen
  \bibfield  {author} {\bibinfo {author} {\bibfnamefont {W.}~\bibnamefont
  {Chen}}, \bibinfo {author} {\bibfnamefont {Z.}~\bibnamefont {Yang}}, \bibinfo
  {author} {\bibfnamefont {Y.}~\bibnamefont {He}}, \bibinfo {author}
  {\bibfnamefont {W.}~\bibnamefont {Ke}}, \bibinfo {author} {\bibfnamefont
  {L.~G.}\ \bibnamefont {Pang}},\ and\ \bibinfo {author} {\bibfnamefont
  {X.-N.}\ \bibnamefont {Wang}},\ }\bibfield  {title} {\bibinfo {title}
  {{Search for the Elusive Jet-Induced Diffusion Wake in $Z/\gamma$-Jets with
  2D Jet Tomography in High-Energy Heavy-Ion Collisions}},\ }\href@noop {}
  {\bibfield  {journal} {\bibinfo  {journal} {Phys. Rev. Lett.}\ }\textbf
  {\bibinfo {volume} {127}},\ \bibinfo {pages} {082301} (\bibinfo {year}
  {2021})}\BibitemShut {NoStop}%
\bibitem [{\citenamefont {Casalderrey-Solana}\ \emph
  {et~al.}(2014)\citenamefont {Casalderrey-Solana}, \citenamefont {Gulhan},
  \citenamefont {Milhano}, \citenamefont {Pablos},\ and\ \citenamefont
  {Rajagopal}}]{bib:Hybrid14}%
  \BibitemOpen
  \bibfield  {author} {\bibinfo {author} {\bibfnamefont {J.}~\bibnamefont
  {Casalderrey-Solana}}, \bibinfo {author} {\bibfnamefont {D.~C.}\ \bibnamefont
  {Gulhan}}, \bibinfo {author} {\bibfnamefont {J.~G.}\ \bibnamefont {Milhano}},
  \bibinfo {author} {\bibfnamefont {D.}~\bibnamefont {Pablos}},\ and\ \bibinfo
  {author} {\bibfnamefont {K.}~\bibnamefont {Rajagopal}},\ }\bibfield  {title}
  {\bibinfo {title} {{A Hybrid Strong/Weak Coupling Approach to Jet
  Quenching}},\ }\href@noop {} {\bibfield  {journal} {\bibinfo  {journal} {J,
  High Energy Phys.}\ }\textbf {\bibinfo {volume} {10}},\ \bibinfo {pages}
  {019} (\bibinfo {year} {2014})},\ \bibinfo {note} {[Erratum: J, High Energy
  Phys. {\bf 09 (2015)}, 175]}\BibitemShut {NoStop}%
\bibitem [{\citenamefont {Adare}\ \emph
  {et~al.}(2010{\natexlab{a}})\citenamefont {Adare} \emph
  {et~al.}}]{bib:PPG106}%
  \BibitemOpen
  \bibfield  {author} {\bibinfo {author} {\bibfnamefont {A.}~\bibnamefont
  {Adare}} \emph {et~al.} (\bibinfo {collaboration} {PHENIX Collaboration}),\
  }\bibfield  {title} {\bibinfo {title} {{Transition in Yield and Azimuthal
  Shape Modification in Dihadron Correlations in Relativistic Heavy Ion
  Collisions}},\ }\href@noop {} {\bibfield  {journal} {\bibinfo  {journal}
  {Phys. Rev. Lett.}\ }\textbf {\bibinfo {volume} {104}},\ \bibinfo {pages}
  {252301} (\bibinfo {year} {2010}{\natexlab{a}})}\BibitemShut {NoStop}%
\bibitem [{\citenamefont {Allen}\ \emph {et~al.}(2003)\citenamefont {Allen}
  \emph {et~al.}}]{bib:PHENIX_inner}%
  \BibitemOpen
  \bibfield  {author} {\bibinfo {author} {\bibfnamefont {M.}~\bibnamefont
  {Allen}} \emph {et~al.} (\bibinfo {collaboration} {PHENIX Collaboration}),\
  }\bibfield  {title} {\bibinfo {title} {{PHENIX inner detectors}},\
  }\href@noop {} {\bibfield  {journal} {\bibinfo  {journal} {Nucl. Instrum.
  Methods Phys. Res., Sec. A}\ }\textbf {\bibinfo {volume} {499}},\ \bibinfo
  {pages} {549} (\bibinfo {year} {2003})}\BibitemShut {NoStop}%
\bibitem [{\citenamefont {Adcox}\ \emph {et~al.}(2003)\citenamefont {Adcox}
  \emph {et~al.}}]{bib:PHENIX_tracking}%
  \BibitemOpen
  \bibfield  {author} {\bibinfo {author} {\bibfnamefont {K.}~\bibnamefont
  {Adcox}} \emph {et~al.} (\bibinfo {collaboration} {PHENIX Collaboration}),\
  }\bibfield  {title} {\bibinfo {title} {{PHENIX central arm tracking
  detectors}},\ }\href@noop {} {\bibfield  {journal} {\bibinfo  {journal}
  {Nucl. Instrum. Methods Phys. Res., Sec. A}\ }\textbf {\bibinfo {volume}
  {499}},\ \bibinfo {pages} {489} (\bibinfo {year} {2003})}\BibitemShut
  {NoStop}%
\bibitem [{\citenamefont {Sickles}\ \emph {et~al.}(2010)\citenamefont
  {Sickles}, \citenamefont {McCumber},\ and\ \citenamefont
  {Adare}}]{bib:AbsNorm}%
  \BibitemOpen
  \bibfield  {author} {\bibinfo {author} {\bibfnamefont {A.}~\bibnamefont
  {Sickles}}, \bibinfo {author} {\bibfnamefont {M.~P.}\ \bibnamefont
  {McCumber}},\ and\ \bibinfo {author} {\bibfnamefont {A.}~\bibnamefont
  {Adare}},\ }\bibfield  {title} {\bibinfo {title} {{Extraction of correlated
  jet pair signals in relativistic heavy ion collisions}},\ }\href@noop {}
  {\bibfield  {journal} {\bibinfo  {journal} {Phys. Rev. C}\ }\textbf {\bibinfo
  {volume} {81}},\ \bibinfo {pages} {014908} (\bibinfo {year}
  {2010})}\BibitemShut {NoStop}%
\bibitem [{\citenamefont {Abelev}\ \emph {et~al.}(2008)\citenamefont {Abelev}
  \emph {et~al.}}]{bib:STARv1_2008}%
  \BibitemOpen
  \bibfield  {author} {\bibinfo {author} {\bibfnamefont {B.~I.}\ \bibnamefont
  {Abelev}} \emph {et~al.} (\bibinfo {collaboration} {STAR Collaboration}),\
  }\bibfield  {title} {\bibinfo {title} {{System-Size Independence of Directed
  Flow Measured at the BNL Relativistic Heavy-Ion Collider}},\ }\href@noop {}
  {\bibfield  {journal} {\bibinfo  {journal} {Phys. Rev. Lett.}\ }\textbf
  {\bibinfo {volume} {101}},\ \bibinfo {pages} {252301} (\bibinfo {year}
  {2008})}\BibitemShut {NoStop}%
\bibitem [{\citenamefont {Adamczyk}\ \emph {et~al.}(2014)\citenamefont
  {Adamczyk} \emph {et~al.}}]{bib:STARv1_2014}%
  \BibitemOpen
  \bibfield  {author} {\bibinfo {author} {\bibfnamefont {L.}~\bibnamefont
  {Adamczyk}} \emph {et~al.} (\bibinfo {collaboration} {STAR Collaboration}),\
  }\bibfield  {title} {\bibinfo {title} {{Beam-Energy Dependence of the
  Directed Flow of Protons, Antiprotons, and Pions in {Au+Au} Collisions}},\
  }\href@noop {} {\bibfield  {journal} {\bibinfo  {journal} {Phys. Rev. Lett.}\
  }\textbf {\bibinfo {volume} {112}},\ \bibinfo {pages} {162301} (\bibinfo
  {year} {2014})}\BibitemShut {NoStop}%
\bibitem [{\citenamefont {Adare}\ \emph {et~al.}(2019)\citenamefont {Adare}
  \emph {et~al.}}]{bib:PPG173}%
  \BibitemOpen
  \bibfield  {author} {\bibinfo {author} {\bibfnamefont {A.}~\bibnamefont
  {Adare}} \emph {et~al.} (\bibinfo {collaboration} {PHENIX Collaboration}),\
  }\bibfield  {title} {\bibinfo {title} {{Measurement of two--particle
  correlations with respect to second-- and third--order event planes in
  {Au+Au} collisions at $\sqrt{{s}_{\mathit{NN}}}=200$ GeV}},\ }\href@noop {}
  {\bibfield  {journal} {\bibinfo  {journal} {Phys. Rev. C}\ }\textbf {\bibinfo
  {volume} {99}},\ \bibinfo {pages} {054903} (\bibinfo {year}
  {2019})}\BibitemShut {NoStop}%
\bibitem [{\citenamefont {Adare}\ \emph
  {et~al.}(2010{\natexlab{b}})\citenamefont {Adare} \emph
  {et~al.}}]{bib:PPG110}%
  \BibitemOpen
  \bibfield  {author} {\bibinfo {author} {\bibfnamefont {A.}~\bibnamefont
  {Adare}} \emph {et~al.} (\bibinfo {collaboration} {PHENIX Collaboration}),\
  }\bibfield  {title} {\bibinfo {title} {{Azimuthal Anisotropy of $\pi^0$
  Production in Au$+$Au Collisions at $\sqrt{s_{NN}}=200$ GeV: Path-Length
  Dependence of Jet Quenching and the Role of Initial Geometry}},\ }\href@noop
  {} {\bibfield  {journal} {\bibinfo  {journal} {Phys. Rev. Lett.}\ }\textbf
  {\bibinfo {volume} {105}},\ \bibinfo {pages} {142301} (\bibinfo {year}
  {2010}{\natexlab{b}})}\BibitemShut {NoStop}%
\bibitem [{\citenamefont {Lacey}\ \emph {et~al.}(2011)\citenamefont {Lacey},
  \citenamefont {Taranenko}, \citenamefont {Ajitanand},\ and\ \citenamefont
  {Alexander}}]{bib:AcousticScaling}%
  \BibitemOpen
  \bibfield  {author} {\bibinfo {author} {\bibfnamefont {R.~A.}\ \bibnamefont
  {Lacey}}, \bibinfo {author} {\bibfnamefont {A.}~\bibnamefont {Taranenko}},
  \bibinfo {author} {\bibfnamefont {N.~N.}\ \bibnamefont {Ajitanand}},\ and\
  \bibinfo {author} {\bibfnamefont {J.~M.}\ \bibnamefont {Alexander}},\
  }\bibfield  {title} {\bibinfo {title} {{Scaling of the higher-order flow
  harmonics: implications for initial-eccentricity models and the 'viscous
  horizon'}}} (\bibinfo {year} {2011}),\ \bibinfo {note}
  {{arXiv:1105.3782}}\BibitemShut {NoStop}%
\bibitem [{\citenamefont {Adam}\ \emph {et~al.}(2016)\citenamefont {Adam} \emph
  {et~al.}}]{bib:alicepi0}%
  \BibitemOpen
  \bibfield  {author} {\bibinfo {author} {\bibfnamefont {J.}~\bibnamefont
  {Adam}} \emph {et~al.} (\bibinfo {collaboration} {ALICE Collaboration}),\
  }\bibfield  {title} {\bibinfo {title} {{Jet-like correlations with neutral
  pion triggers in $pp$ and central Pb-Pb collisions at 2.76 TeV}},\ }\href
  {https://doi.org/10.1016/j.physletb.2016.10.048} {\bibfield  {journal}
  {\bibinfo  {journal} {Phys. Lett. B}\ }\textbf {\bibinfo {volume} {763}},\
  \bibinfo {pages} {238} (\bibinfo {year} {2016})}\BibitemShut {NoStop}%
\bibitem [{\citenamefont {Aad}\ \emph {et~al.}(2019)\citenamefont {Aad} \emph
  {et~al.}}]{bib:AtlasRadialModification}%
  \BibitemOpen
  \bibfield  {author} {\bibinfo {author} {\bibfnamefont {G.}~\bibnamefont
  {Aad}} \emph {et~al.} (\bibinfo {collaboration} {ATLAS Collaboration}),\
  }\bibfield  {title} {\bibinfo {title} {{Measurement of angular and momentum
  distributions of charged particles within and around jets in Pb$+$Pb and $pp$
  collisions at $\sqrt{s_{NN}}=5.02$ TeV with the ATLAS detector}},\
  }\href@noop {} {\bibfield  {journal} {\bibinfo  {journal} {Phys. Rev. C}\
  }\textbf {\bibinfo {volume} {100}},\ \bibinfo {pages} {064901} (\bibinfo
  {year} {2019})}\BibitemShut {NoStop}%
\end{thebibliography}

 
\end{document}